\documentstyle[preprint,aps]{revtex}
\begin{document}
\draft
\title{Single Particle Hopping
Between Luttinger Liquids: A Spectral Function Approach}
\author{David G. Clarke}

\address{IRC in Superconductivity and Cavendish Laboratory,
University of Cambridge,\\
Cambridge, CB3 0HE, United Kingdom \\}
\author{S. P. Strong}
\address{
NEC Research Institute, 4 Independence Way,
Princeton, NJ, 08540, U.S.A.\\}
\date{June 21, 1995}
\maketitle
\begin{abstract}
We present a pedagogical account of our approach to the problem of
Luttinger liquids coupled by interliquid single particle hopping.
It is shown that the key issue is that of coherence/incoherence of
interliquid hopping, and not of relevance/irrelevance in a renormalization
group
sense. A clear signal of coherence, present in the case of coupled Fermi
liquids, is absent for Luttinger liquids, and we argue for the existence
of an incoherent regime when the interliquid hopping rate is sufficiently
small.
The problem is relevant to any sufficiently anisotropic, strongly
correlated
metal, and in particular to understanding
the anomalous c-axis conductivity in the cuprate superconductors,
and the physics of the quasi-1D organic conductors.

\end{abstract}
\pacs{PACS numbers: 71.27+a, 72.10-d, 72.10Bg, 74.25-q}

\narrowtext

\section{Introduction}

The development of a complete theoretical understanding of the high-temperature
superconducting cuprates (HTSC's) represents one of the most challenging tasks
presently facing the field of condensed matter physics. Apart from the
extraordinarily
high superconducting transition temperatures, $T_c$, these materials exhibit a
bewildering
array of anomalous normal state properties. It is clear that the normal state
is not
a (Landau) Fermi liquid: on this point there is practically universal
agreement.
On the other hand, the {\em mechanism\/} giving rise to the non-Fermi liquid
normal
state is still controversial. Nevertheless, it would seem perverse to reject
the
notion that
there is an intimate connection between
high $T_c$ and the fact that the normal state is an anomalous metal.

For a long time, Anderson has emphasized another aspect of the experimental
data for the normal state. This is the {\em qualitative\/} difference
between the in-plane (ab-plane) and inter-plane (c-axis) physics. Indeed,
the very use of the term `in-plane' presupposes such a qualitative
difference: the c-axis transport is {\em not\/} simply
the same as the ab-plane transport up to a scaling factor to account for
anisotropy. This is an important point, for recent studies \cite{ruthenate}
on Sr$_2$RuO$_4$, a structural analogue of  La$_2$CuO$_4$, show that the
low-temperature dc-resistivity has the {\em same\/} temperature dependence
in both the ab-plane and c-axis directions, with an anisotropy of up to
$\approx 500$. It is no coincidence that this temperature dependence is
Fermi-liquid like.

An extensive account of c-axis experiments in the cuprates has been given
by Cooper and Gray \cite{cooper_gray}. As far as the c-axis conductivity,
$\sigma_c(\omega)$, is concerned, the key observations for
superconducting samples are a dc-conductivity of order of or
well below the Mott limit of minimum metallic conductivity, and
a frequency dependence which is completely incoherent in the sense of
the {\em absence\/} of a Drude term
\cite{cooper_gray,uchida,cooper}.
It has been argued that YBa$_2$Cu$_3$O$_7$ has a Drude-like component
\cite{cooper,schutzmann}, but the width is anomalously large.

The underlying idea of Anderson's ``confinement'' hypothesis \cite{phil_prl}
toward
understanding the strange behavior
of the c-axis conductivity is to associate the observed incoherent
conductivity with the non-Fermi liquid ab-plane properties.
{}From this point
of view, it is significant that in overdoped samples of La$_{2-x}$Sr$_x$CuO$_4$
the appearance of a Drude-like term in $\sigma_c(\omega)$ coincides with a
crossover to Fermi liquid-like ab-plane properties \cite{uchida}.

In Bi$_2$Sr$_2$CaCu$_2$O$_8$, this picture of incoherent c-axis transport
is confirmed by the photoemission experiments of Ding {\em et al.\/}
\cite{ding}.
These authors observe a {\em single\/} Fermi surface, up to a resolution of
$\sim 10 meV$. This is in strong disagreement with the band theory prediction
of two Fermi surfaces split by an energy of $\sim 400 meV$, this energy
splitting
being the direct result of coherent single particle hopping between adjacent
CuO$_2$ planes.

The issue we address in this paper is relevant to {\em any\/} sufficiently
anisotropic, strongly correlated metal, and therefore in particular to
the quasi-1D metals. We have already proposed one application of our
ideas to partially understanding the anomalous magnetoresistance in the
quasi-1D organic conductor (TMTSF)$_2$PF$_6$ \cite{magic}. An extensive
discussion of this and of more recent experiments supporting our
point of view \cite{danner_chaikin} will appear elsewhere \cite{long}.

Given the acceptance of a non-Fermi liquid (NFL) state in the planes,
the experimental observations in the cuprates force us to carefully
examine the problem of 2D NFL's coupled by single particle
hopping terms which transfer  {\em real\/} electrons from one liquid
to a physically adjacent one. The Hamiltonian we consider is simply
\begin{equation}
H=\sum_{i} H_{{\rm NFL}}^{(i)}+t_{\perp}\sum_{i,x}
\{c_{i,\sigma}^{\dag}(x)c_{i+1,\sigma}(x) +{\rm h.c.}\}
\end{equation}
{\em i.e.\/}, $t_{\perp}$ is
local in real space, and thus $k$-diagonal in momentum space.
Of great import is the realization that real electrons are {\em not\/}
eigenstates of the NFL: there are no electron-like quasiparticles, {\em
i.e.\/},
$Z=0$. As such, removing an electron from, or adding an electron to, one of the
liquids will necessarily ``disturb'' the liquid in some sense. The effect is
similar to the way in which the flipping of a Kondo spin causes an
orthogonal rearrangement of the Fermi sea with which it is in contact.

It is also important to realize that the central issue of concern here
is {\em not\/} one of relevance or irrelevance of $t_{\perp}$ in the
renormalization
group (RG) sense. In fact, the existence of a sharp Fermi surface is
sufficient evidence of itself to point to the relevance of $t_{\perp}$.
Within a Luttinger liquid framework, the singularity of $n(k)$ near $k_F$
behaves as
\begin{equation}
\label{n(k)}
\frac{\partial n(k)}{\partial k}\sim -| k-k_F|^{2\alpha-1}
\end{equation}
At the bare level of the RG $t_{\perp}$ is irrelevant iff $2\alpha > 1$,
in which case there would not be a sharp Fermi surface at all! A sharp Fermi
surface therefore implies the relevance of $t_{\perp}$.
Nevertheless,
the relevance of $t_{\perp}$ in the RG sense does {\em not\/} guarantee that
single particle interliquid hopping will be {\em coherent\/}.
An
example is afforded by the problem of a two-level system (TLS)
coupled to a dissipative bath. Here it is known that there are situations
where the tunneling operator between the two (usually degenerate) states
is relevant, yet nevertheless tunneling is completely incoherent.

Before briefly discussing the TLS problem, it is helpful to begin with some
very simple, yet instructive, considerations.

\section{Fermi's ``Golden Rule''}

Consider the textbook derivation of the so-called ``Golden Rule'' of Fermi:
a particle is in state $| i\rangle$, an eigenstate for all times $t<0$.
At time $t=0$ a perturbation $V$ is turned on. To $O(V^2)$ the probability
for the particle to be observed in state $| f\rangle$ at time $t>0$ is
\begin{eqnarray}
&&\left|\frac{-i}{\hbar}\int_0^t dt'\langle f| V| i\rangle
e^{i/\hbar(E_f-E_i)t'}\right|^2 \nonumber\\
&=& | V_{if}| ^2\left|\frac{e^{i\Delta E t/\hbar}-1}{\Delta E}
\right| ^2 \nonumber\\
&=& 4\frac{| V_{if}| ^2}{(\Delta E)^2}\sin^2\left(\frac{\Delta E\;t}
{2\hbar}\right) \nonumber
\end{eqnarray}
Assuming $V_{if}$ depends only on $\Delta E$, the probability to find
the particle in its original state $| i\rangle$
at time $t>0$ is then just
\begin{eqnarray}
P(t)&=&1-4\sum_{f}\frac{| V_{if}| ^2}{(E_f-E_i)^2}
\sin^2\left(\frac{(E_f-E_i)t}{2\hbar}\right) + O(V^4) \nonumber\\
&\rightarrow&1-4\int d\epsilon| V(\epsilon)| ^2\rho(\epsilon)
\frac{\sin^2(\epsilon t/2\hbar)}{\epsilon^2}
\end{eqnarray}
where $\rho(\epsilon)$ is the density of final states of energy
$\epsilon$ relative to $E_i$. If it is now assumed that
$| V(\epsilon)| ^2\rho(\epsilon)\approx V^2\rho_0$, a constant, over
some interval $[-\Lambda,\Lambda]$ of $\epsilon$, then
\begin{eqnarray}
1-P(t)&\approx&4V^2\rho_0\int_{-\Lambda}^{\Lambda}
 d\epsilon
\frac{\sin^2(\epsilon t/2\hbar)}{\epsilon^2} \nonumber\\
&\stackrel{\Lambda t/2\hbar\gg 1}{\longrightarrow}&
4V^2\rho_0\int_{-\infty}^{\infty}
 d\epsilon
\frac{\sin^2(\epsilon t/2\hbar)}{\epsilon^2} \nonumber\\
&=&\frac{2\pi}{\hbar}(V^2\rho_0)t
\end{eqnarray}
Thus, to lowest order in the perturbation, the rate at which the particle
leaves $| i\rangle$ is given by
\[
\Gamma=\frac{2\pi}{\hbar}(V^2\rho_0)
\]
which is Fermi's Golden Rule.

In many applications of the Golden Rule it is assumed, usually without proof,
that this lowest order result can be extended to all $t$ via
exponentiation to give $P(t)=e^{-\Gamma t}$.

There are two crucial steps in the derivation of the Golden rule. The
first is the replacement of $| V(\epsilon)| ^2\rho(\epsilon)$ by a constant
$V^2\rho_0$. This is a reasonable approximation as long as
$| V(\epsilon)| ^2\rho(\epsilon)$ is nonsingular in the interval
$[-\Lambda,\Lambda]$. The second is the assumption that $\Lambda t/2\hbar\gg
1$.
This requires the spectral weight to be spread out over an energy scale larger
than
$\hbar/t$.

A trivial example of a situation where the Golden Rule is invalid is the case
where there is just {\em one\/} final state, $| f\rangle$. Then
\begin{eqnarray}
P(t)&=&1-\frac{| V_{if}| ^2}{(E_f-E_i)^2}
\sin^2\left(\frac{(E_f-E_i)t}{2\hbar}\right) \nonumber\\
&\stackrel{E_f\rightarrow E_i}{\longrightarrow}&1-\frac{| V_{if}| ^2
t^2}{\hbar^2}
\end{eqnarray}
Such a result precludes any interpretation in terms of a decay rate. Indeed,
this trivial problem is solved exactly by diagonalization of the Hamiltonian
\[
H=\left(\begin{array}{cc} E_i & V^* \\
	V & E_f
	\end{array}\right)
\]
The oscillation frequency of $P(t)$ is
\[
\hbar\omega_{osc}=[(\Delta E)^2+4|V|^2]^{1/2}
\]
Note that the $O(V^2)$ perturbative calculation does not directly give this
oscillation frequency. Care must be taken in interpretation even in the case
$4|V|^2/(E_f-E_i)^2\ll 1$. Here, the perturbation expansion appears controlled,
$P(t)$ oscillating between 1 and $1-4|V|^2/(E_f-E_i)^2$ with frequency
$(E_f-E_i)/\hbar$, yet the exact solution shows that the true oscillation
frequency is $\omega_{osc}$. The $O(V^2)$ calculation fails to pick up the
frequency shift. The oscillation of $P(t)$ is most striking in the {\em
degenerate\/}
case $E_f=E_i$. In this case the $O(V^2)$ calculation picks up the first term
in
an expansion of the exact result
\begin{equation}
P(t)=\cos^2(|V|t/\hbar)=\frac{1}{2}\left\{\cos\left(\frac{2|V|t}{\hbar}\right)+1
\right\}
\end{equation}
The particle oscillates between $| i\rangle$ and $| f\rangle$ with frequency
$2|V|/\hbar$. Such spectacular oscillation effects occur in the famous
($K_L$, $K_S$) system in particle physics, and were at one time
proposed as a possible resolution of the solar neutrino problem. The effect is
the hallmark of a truly ``quantum'' system and is at the heart of measurement
theory and the kinds of {\em gedanken\/}  problems which began with
Schr\"{o}dinger's  worries about cats.

The situation where there is just a single final state corresponds to
the spectral density $\rho(\omega)=\delta(\omega)$. One can consider other
types of singular densities of states, perhaps the simplest being those of
``edge'' type,
\[
\rho(\epsilon)=(1-\gamma)\Lambda^{\gamma
-1}\theta_+(\epsilon)\epsilon^{-\gamma}
\]
($\gamma <1$). To $O(V^2)$ (assuming $V\approx\rm{constant}$)
\begin{eqnarray}
1-P(t)&=&4V^2\int_0^{\Lambda}d\epsilon\;\epsilon^{-(\gamma+2)}
\sin^2(\epsilon t/2\hbar) \nonumber\\
&\stackrel{\Lambda t\gg\hbar}{\longrightarrow}&
\Lambda^{\gamma-1}4V^2\left(\frac{t}{2\hbar}\right)^{1+\gamma}
\int_0^{\infty}dx\;\frac{\sin^2 x}{x^{2+\gamma}}
\end{eqnarray}
For $\gamma < -1$, $P(t)\rightarrow 1$ for all $t$ in the
limit $V\rightarrow 0$: the particle
is localized in the state $| i\rangle$, and $V$ is an irrelevant perturbation.
$\gamma\rightarrow 0$ corresponds to a uniform density of states, in which case
the Golden Rule is applicable, while $\gamma\rightarrow 1$ corresponds to the
extreme coherence limit in which the Golden Rule fails. A natural question to
ask
is: what is the true nature of $P(t)$ in the intermediate cases $-1 < \gamma <
1$?
Such is the question studied in the context of a more particular problem, that
of
a two-level system (TLS) in contact with a dissipative bath.

\section{A Summary of The Two-Level System Problem}

Following Ref. \cite{TLS}, we define
the two level system
model by the Hamiltonian:
\begin{eqnarray}
\label{eq:TLS_ham}
H_{\rm TLS} & = &  \frac{1}{2} \Delta \sigma_x + \frac{1}{2}
\epsilon \sigma_z + \sum_i  \left\{\frac{1}{2} m_i \omega_i
x_i^2  + \frac{1}{2} p_i^2 / m_i \right\} \\
& &+ \frac{1}{2}  \sigma_z \sum_i C_i x_i
\nonumber
\end{eqnarray}
Here $C_i$ is the coupling to the $i$th oscillator,
and $m_i$, $\omega_i$, $x_i$ and $p_i$
are the mass, frequency, position
and momentum of the $i$th oscillator, respectively.

The model describes a single quantum mechanical
degree of freedom which can be in either of two states
and which is coupled to a bath of harmonic oscillators.
The $\sigma_i$ are Pauli matrices, so
we will refer to the two-state degree of
freedom as a `spin' for convenience.
We are primarily interested in the $\epsilon =0$
case and subsequent discussion refers to this case
unless otherwise stated.  In this case, one may
think of the model as one in which a particle
tunnels at a rate $\Delta/2$ between two degenerate states (labelled by
$\sigma_z=\pm 1$).  The environment, represented by the bath of
oscillators,
influences the tunneling
because the bath
is sensitive to which of the states the spin
is in.

The TLS model provides the prototypical example
of a quantum to classical crossover. For
$C_i =0$ the model represents the quantum mechanics
of an isolated two state system, whereas for
sufficiently strong coupling to the environment
the dynamics of the spin,
if followed without reference
to the oscillator bath, are dissipative and
no quantum coherence effects are observable \cite{TLS}.
In fact, this is how one generally expects
classical behavior to emerge for macroscopic systems:
the macroscopic degrees of freedom exchange energy with
an enormous number of unobserved microscopic degrees
of freedom and are therefore unable to maintain
a definite phase long enough for
quantum interference effects to manifest themselves.

Consider first the case
where the coupling of the spin to the
environment vanishes, i.e. $\alpha=0$, and where the
spin is prepared in an eigenstate of $\sigma_z$.
The exact eigenstates of the spin are
the $\sigma_x$ eigenstates which are split by an energy
$\Delta$. The initial state of the system is therefore
a superposition of these two states with a definite
phase between them.
Since the two states have different energy, this phase
is not time independent. For
vanishing coupling to the environment the phase remains
well defined indefinitely. The time dependence of the phase
therefore results in observable oscillations in
the expectation value of
$\sigma_z$, in fact
(in units where $\hbar = 1$)
$\langle\sigma_z(t)\rangle = \cos  \Delta t$.
Such oscillations are a quantum interference effect. In general,
we would expect such
oscillations to also occur when the spin is coupled to the environment
provided the spin is capable of
flipping without exchanging an amount of energy with
its environment sufficient for the randomization of the phases of
the various states in the superposition.
Indeed \cite{TLS}, in the TLS model
the oscillations persist for a range of couplings
to the environment,
albeit with coupling dependent
damping of the oscillations. Such damping results
from the exchange of energy between the spin and the environment
during the course of a typical flip.

In order to study the quantum oscillations, or lack thereof,
in a TLS, it is natural, in the light of the discussion above, to
invoke the following prescription. First, the
system is prepared by clamping the
spin into the $\sigma_z =1$ state for all $t<0$,
allowing the oscillator bath to relax to equilibrium.
The spin is then released at $t=0$, and one attempts to determine
$\langle\sigma_z(t)\rangle$ for positive times, $t$.
This is, quite generally, an appropriate quantity to study for
questions about macroscopic quantum coherence,
for if the
spin represents a generic macroscopic quantum
degree of freedom which the experimenter can observe
and control, whereas the oscillators represent
microscopic degrees of freedom which are beyond
both control and observational capacities of the
experimenter, then the above ``clamping'' prescription is exactly the sort of
preparation which
is possible experimentally.
Calculation of  $\langle\sigma_z(t)\rangle$ is equivalent to
determining $P(t)$, the probability of observing the spin
in the $\sigma_z =1$ state at time $t>0$ \cite{fn_P(t)}. The
two are simply connected by $\langle\sigma_z(t)\rangle=2P(t)-1$.
The signature of quantum coherence in
$\langle\sigma_z(t)\rangle$ will be the presence of oscillations
(damped or otherwise) in contrast to the incoherent
relaxation ($\langle\sigma_z(t)\rangle \sim
e^{- \Gamma t}$)
which must result if the spin exchanges
sufficient energy with the environment to
randomize its phase in tunneling between the two
$\sigma_z$ eigenstates. The transition to
purely incoherent relaxation  is indeed found to occur
in the TLS problem even at short times \cite{TLS}
when $\alpha > \frac{1}{2}$.

Within this formulation of the TLS problem
it is convenient to make the canonical transformation
\begin{equation}
H_{\rm TLS}^{\prime} = \hat{U} H_{\rm TLS} \hat{U}^{-1}
\end{equation}
where
\begin{equation}
\label{eq:xform}
\hat{U} = \exp\left\{-\frac{1}{2} \sigma_z \sum_i \frac{C_i}{m_i \omega_i^2}
\hat{p}_i\right\}
\end{equation}
$\hat{p}_i$ is the momentum operator of the $i$th oscillator.
The new Hamiltonian takes the form
\begin{equation}
\label{eq:TLS_ham_can}
H_{\rm TLS}^{\prime} = \frac{1}{2} \Delta(\sigma^+ e^{-i \Omega} + {\rm h.c.})
+
H_{{\rm oscillators}}
\end{equation}
where $\Omega = \sum_i \frac{C_i}{m_i \omega_i^2} p_i$.
This transformation is very helpful in guiding one's intuition
for it has removed all coupling of the spin to the oscillator bath.
One can now introduce oscillator creation and annihilation operators,
$a_i^{\dag}$, $a_i$, and in so doing one observes that the price paid
for removing all coupling of the spin to the bath is that the tunneling
operator between the
two states has been replaced by an operator which creates and
destroys excitations of the bath as well as changing
the state of the spin. It is then clear that quantum oscillations
will be in danger of being wiped out should the low-frequency
oscillator density of states and/or the low-frequency couplings $C_i$
be sufficiently large.

For our purposes, we shall only consider the
so-called {\em ohmic regime\/} \cite{TLS} where the
density of states of the bath, and its couplings, $C_i$, to the
spin, are such that the
``spectral density'', $J(\omega)$, of
the bath is given by
\begin{eqnarray}
\label{eq:ohmic}
J(\omega) &\equiv&\frac{\pi}{2} \sum_i \frac{C_i}{m_i \omega_i}
\delta(\omega-\omega_i) \\
\nonumber
 & = & 2 \pi~ \alpha~ \omega \exp(-\omega/\omega_c)
\end{eqnarray}
$\alpha$ is a positive constant measuring the strength
of the coupling to the bath and $\omega_c$ is an ultraviolet
cutoff.

In the ohmic regime
the two point correlation function of $\sigma^+ e^{-i \Omega}$
is
\begin{eqnarray}
\label{eq:TLScorr}
\langle \sigma^+ e^{-i \Omega(t)} \sigma^- e^{i \Omega(0)}\rangle
&=&\exp \left\{-\int_0^{\infty}
\frac{1 - e^{-i\omega t}}{\omega^2}
J(\omega)\right\}\nonumber\\
& = &
\exp \left\{-2  \alpha \int_0^{\infty}
\frac{1 - e^{-i\omega t}}{\omega}
e^{-\omega/\omega_c} \right\}\\
\nonumber & \sim &
e^{i \pi \alpha} (\omega_c t)^{-2 \alpha}
\end{eqnarray}
{}From the correlation function we can immediately construct the
spectral function of the operator $ e^{i \Omega}$
in the low energy, universal regime:
\begin{eqnarray}
\label{eq:TLSrho}
\rho_{\Omega}(\omega) & = & \sum_m |\langle m | e^{i \Omega} | GS \rangle |^2
\delta(\omega - E_m) \\
\nonumber & = &
\Gamma^{-1}(2 \alpha)~\theta_+(\omega)~
\omega^{-1+2 \alpha} \omega_c^{-2 \alpha}
e^{-\omega/\omega_c}
\end{eqnarray}
where $\{m\}$ is a complete set of oscillator eigenstates with
energies $E_m$ and $|GS\rangle$ is the oscillator ground state.
The spectral function is  of ``edge'' type, and the physics is
largely determined by whether or not the edge singularity is
divergent as $\omega\rightarrow 0$.

We may now construct the short time approximation to $P(t)$
by using
the spectral function above and ordinary time dependent
perturbation theory.  We find
\begin{equation}
\label{eq:Pt1}
P(t) = 1 -  \frac{\Delta^2}{2}  \int d\omega \rho_{\Omega}(\omega)
\frac{\sin^2(\omega t)}{\omega^2} + \cdots
\end{equation}
Notice that when $\alpha > 1$,
$\rho_{\Omega}(\omega) \sim \omega^{-1+2 \alpha}$ results in
an infrared convergent $P(t)$. In the limit $\Delta \rightarrow
0$, $P(t) \rightarrow 1$ for all $t$. This corresponds to
the irrelevance of $\Delta$ and to localization of the
spin.

For $\alpha \rightarrow 0$, $\rho_{\Omega}(\omega)\rightarrow\delta(\omega)$
and $P(t)\rightarrow 1 - \frac{\Delta^2}{2} t^2 + ...$, in agreement with the
expansion of the exact $\alpha=0$ result $P(t) = \cos^2(\Delta t/2)$ or,
equivalently, of $\langle\sigma_z(t)\rangle=\cos(\Delta(t))$.
For $0 < \alpha < 1$ we are in a more complicated region.
Clearly the difference between $P(t)$  and $1$ grows to
order unity for any arbitrarily small $\Delta$ throughout
this region (this simply reflects the renormalization
group relevance of $\Delta$) and one might at first sight
be tempted to conclude that
$P(t)$ would undergo damped oscillations with
a period approximately given by the time at which
$\Delta^2 \int d\omega \rho_{\Omega}(\omega)
\frac{\sin^2(\omega t)}{\omega^2} \sim 1$.
However, for
$\alpha = \frac{1}{2}$, the spectral function is flat and
featureless out to the cutoff scale.  A flat spectral
function is exactly the condition under which
the Golden Rule approximation
should be valid, implying incoherent decay without
any recurrence effects or oscillations.

We may scale out the time dependence in (\ref{eq:Pt1}) to obtain to
$O(\Delta^2)$ and for $\omega_c t\gg 1$
\begin{equation}
P(t) \approx 1 -  \alpha\Delta^2~\omega_c^{-2\alpha}
t^{2-2\alpha}  \int_0^{\infty} dx
\frac{\sin^2x}{x^{3-2\alpha}}
\end{equation}
(for simplicity, we  have replaced the cutoff $e^{-\omega/\omega_c}$ by
a hard cutoff at $\omega_c$).
Thus,
for $\alpha > \frac{1}{2}$, where the spectral function
for the tunneling operator is vanishing at low frequencies,
we see that the $O(\Delta^2)$ term in $P(t)$ grows even {\em slower\/}
than $t$, suggesting an even ``more incoherent'' decay of $P(t)$.
If we define $\Gamma(t)=-dP(t)/dt$, the rate at which the spin flips,
then in this regime $\Gamma(t)$ is bounded for all $t$. For the special
value $\alpha=1/2$, $\Gamma(t)=\Gamma$, a constant, and a naive
re-exponentiation
of the Golden Rule is $P(t)=(1+e^{-\Gamma t})/2$, corresponding to
$\langle\sigma_z(t)\rangle=e^{-\Gamma t}$. For $\alpha > 1/2$ it would
appear reasonable to expect exponential relaxation, too. A self-consistent
approximation to determining the relaxation rate $\Gamma$ for small
$\Delta/\omega_c$ involves cutting off the $\omega$-integral at
$\omega\sim\Gamma$
to give $\Gamma\sim\Delta^2\omega_c^{-2\alpha}\Gamma^{2\alpha-1}$ yielding
\[
\Gamma\approx\Delta\left(\frac{\Delta}{\omega_c}\right)^{\alpha/(1-\alpha)}
\]
The right hand side is in fact nothing but $\Delta_{\rm ren}$, the renormalized
tunneling rate which emerges from an RG analysis.

The true behavior of $P(t)$ in the region $1/2 < \alpha <1$ is actually not
rigorously known \cite{TLS}, but there are reasons for believing that the
self-consistent argument given above is not too far from the truth. The
true decay of $P(t)$ is probably not simply exponential relaxation, but the
key point is that there are not any oscillations. Thus, despite the fact that
the RG approach yields the same scale $\Delta_{\rm ren}$ as the self-consistent
approach, it fails to distinguish between an essentially {\em coherent\/}
$\Delta_{\rm ren}$ and a completely {\em incoherent\/} one.

The important
physical
effect of finite $\alpha$ is that there is a substantial contribution
to $P(t)$ from transitions to states with energies that are
larger than the putative renormalized oscillation frequency.
When the amount of weight in these transitions is larger
than the amount of weight in transitions to low energy
states it no longer makes sense to consider the effects
of $\Delta$ to be coherent. Effectively, each change of state
is accompanied by the creation or annihilation of sufficient
numbers of bosons in the environmental bath that the
phase of the spin is randomized.
Intuitively, one crosses over
from degenerate or nearly degenerate perturbation theory to
non-degenerate perturbation theory (as opposed to the
transition to irrelevant $\Delta$ where the long time
perturbation theory becomes convergent).

For $0<\alpha<1/2$, $\Gamma(t)$ is unbounded and any attempt at
characterizing $P(t)$ by exponential relaxation fails, as indeed
it must as $\alpha\rightarrow 0$. It is believed \cite{TLS} that
the true behavior of $P(t)$ in this region of $\alpha$ is a damped
oscillation with oscillation frequency
$\omega_{\rm osc}=\cos(\pi\alpha/(2-2\alpha))$ and damping
$\Gamma=\sin(\pi\alpha/(2-2\alpha))$, plus an incoherent background.

In conclusion, the key point we wish to make here is that in the TLS
model the qualitative
behavior of $P(t)$, in the sense of whether or not it exhibits oscillations,
{\em i.e.\/} quantum coherence, can actually be determined
from lowest order perturbation
theory. The special point $\alpha=1/2$, at which the Golden Rule is
believed applicable, separates the region of completely incoherent behavior,
$1/2\leq\alpha <1$, from that of damped oscillations, $0<\alpha<1/2$.
In the former region, an ``extended Golden Rule'' ({\em i.e.\/}
the self-consistency argument above) works and $P(t)$ exhibits incoherent
behavior.
In the latter, the extended Golden Rule fails and, in fact,
$P(t)$ exhibits (damped) oscillations.

\section{Coupled Luttinger Liquids}

With this preparation, we now turn to the problem of interest, that
of NFL's coupled by interliquid electron hopping operators. There are several
issues at stake here, of varying complexity, but the central one is whether
or not interliquid single particle hopping is coherent in the limit of small
$t_{\perp}$. If it were, one would have to then address some other questions;
for example, would the resulting ground state be a NFL in one higher dimension
with a warped Fermi surface, or would it be a Fermi liquid?

On the other hand, if the interliquid hopping is incoherent, there cannot
be any coherent interliquid velocity nor any warping of the Fermi surface.
There will be dramatic implications for interliquid properties. In particular,
the interliquid conductivity will not exhibit a Drude term.

In order to be able to make precise calculations,
we restrict ourselves
to the case of coupled 1D liquids, which are {\em Luttinger liquids\/} (LL), a
specific type of NFL. To the extent that the Luttinger liquid concept
can be extended to two dimensions,
we expect
our results
to be generalizable.

Ideally, we would like to tackle the problem of $N$ coupled liquids, for
$N\rightarrow\infty$. However, the problem of two coupled liquids ought to
be sufficient to settle the coherence issue. In any case,
for the purposes of the calculation
it makes no difference if we restrict ourselves to just two liquids
since we calculate
only to $O(t_{\perp}^2)$.

It is a non-trivial matter to determine {\em how\/} one should go about
settling the coherence/incoherence issue. Our approach utilizes the similarity
to the TLS problem with ohmic dissipation. We have
\begin{equation}
H= H_{{\rm LL}}^{(1)}+H_{{\rm LL}}^{(2)}+t_{\perp}\sum_{x}
\{c_{\sigma}^{(1)\dag}(x)c_{\sigma}^{(2)}(x) +{\rm h.c.}\}
\end{equation}
The connection to the TLS-type physics is made by first bosonizing the
Luttinger liquids, which then play the role of two baths of spin and charge
bosons. Under the bosonization the interliquid hopping operators become
exponentials of spin and charge boson creation and annihilation operators.
They resemble the operators $e^{\pm i\Omega}$ of the TLS Hamiltonian,
$H_{\rm TLS}^{\prime}$. The $t_{\perp}$ operator acts to raise
the particle number of one chain by 1, and lower the other by 1, analogous
to the action of the spin flip operators in the TLS.
Moreover, the $t_{\perp}$ operator has a power law two-point function.
Thus, it is very similar to the tunneling operator $(\sigma^+e^{-i\Omega}
+\sigma^-e^{i\Omega})$ in the ohmic regime of the TLS, and
$H_{{\rm LL}}^{(1)}+H_{{\rm LL}}^{(2)}$ plays a role similar to the
oscillator bath in the TLS.

Despite the striking similarity, however, there is no precise mapping
of $H$ to $H_{\rm TLS}$ for the simple reason that in the TLS problem
there is just a {\em single\/} tunneling particle, and this particle
is {\em distinct\/} from the oscillator bath, while in the coupled LL
problem there are $N$ particles which can hop from liquid to liquid and,
moreover, these particles are themselves the {\em source\/} of the
dissipative bath. The most natural variable analogous to $\sigma_z$
is $\Delta N=N_2-N_1$, the particle number difference between the two
liquids, with the obvious difference that $\Delta N$ is not simply
two-valued.

However, despite these problems with the analogy, the function $P(t)$
in the TLS {\em is\/} readily generalized to the coupled LL problem.
For the LL problem we define
\begin{equation}
P(t)\equiv|\langle O_1O_2| e^{iH_0 t}e^{-iHt}| O_1O_2\rangle|^2
\end{equation}
Here $| O_1O_2\rangle$ denotes the product of the ground states
$| O_1\rangle$, $| O_2\rangle$ of each Luttinger liquid in the
absence of $t_{\perp}$. For $t<0$, $H= H_{{\rm LL}}^{(1)}+H_{{\rm LL}}^{(2)}$,
and at $t=0$ the interliquid hopping is turned on. The particle number
difference $\Delta N$ entails a Fermi momentum difference $\Delta k$
and a chemical potential difference $\Delta\mu$.


Suppose that, instead of being Luttinger liquids, the 1D liquids
were free Fermi gasses. The the Hamiltonian becomes a direct product
$H=\otimes_k H_k$, where
\[
H=\left(\begin{array}{cc} E_k & t_{\perp} \\
	t_{\perp} & E_k
	\end{array}\right)
\]
so that
\[
P(t)=\cos^{2\Delta N}(t_{\perp}t)
\]
Perturbation theory picks up the $O(t_{\perp}^2)$ term correctly,
\[
P(t)\sim 1-\Delta N t_{\perp}^2 t^2
\]
This is precisely the type of behavior for which Golden Rule
or extended Golden Rule, {\em i.e.\/} incoherent, type methods
fail and for a very clear reason: quantum coherence is established
separately for each $k$ in a very trivial way.

When interactions between electrons within a given liquid are included,
$H$ can no longer be written in this direct product form. We might
suspect, however, that for true (Landau) Fermi liquids, where the
Landau quasiparticle concept is valid, an approximate decomposition
into a direct product of quasiparticle Hamiltonians would be possible.
On the other hand, the situation for coupled Luttinger liquids, where the
quasiparticle concept completely breaks down, is not at all obvious.

In what follows, we outline an approach to calculating $P(t)$ using
spectral functions (complete details will be given elsewhere \cite{long}).
This method is perhaps more illuminating than the space-time Green's
function method used by us previously \cite{prl}. The ``shape'' of the spectral
function which determines $P(t)$ can be examined to determine the nature
of the interliquid hopping processes.

To $O(t_{\perp}^2)$ we have
\begin{equation}
1-P(t)=2t_{\perp}^2L{\rm Re}\int_0^t dt_1 \int_0^{t_1} dt_2
\int dx\left\{\langle c^{(1)}(x,t_1)c^{(1)\dag}(0,t_2)\rangle
\langle c^{(2)\dag}(x,t_1)c^{(2)}(0,t_2)\rangle +(1\leftrightarrow 2)
\right\}
\end{equation}
where the superscripts on the electron operators label the chain in
which the operator acts. For convenience, we again define
$\Gamma(t)\equiv -dP(t)/dt$ which can be written in a spectral function
form as
\begin{equation}
\Gamma(t)=2t_{\perp}^2L\int\frac{d\omega}{2\pi}
\frac{\sin\omega t}{\omega}\{A_{12}(\omega)+A_{21}(\omega)\}
\end{equation}
where
\begin{equation}
A_{ij}(\omega)=\int\frac{d\omega'}{2\pi}
\int\frac{dk}{2\pi}
{\cal J}_1^{(i)}(k,\omega'){\cal J}_2^{(j)}(k,\omega'-\omega)
\end{equation}
and ${\cal J}_{1,2}(k,\omega)$ are the Fourier transforms of
\begin{eqnarray*}
{\cal J}_1(k,t)&\equiv&\langle c(k,t)c^{\dag}(k,0)\rangle\\
{\cal J}_2(k,t)&\equiv&\langle c^{\dag}(k,0)c(k,t)\rangle
\end{eqnarray*}
In the $T=0$ limit,
\begin{equation}
{\cal J}_{1,2}(k,\omega')=
\theta_{\pm}(\omega'-\mu)\rho(k,\omega'-\mu)
\end{equation}
where $\rho(k,\omega)$ is the electron spectral function as conventionally
defined.

Physically, $A_{12}(\omega)$ is the effective spectral function governing hops
in which an electron hops {\em to\/} liquid 1, {\em from\/} liquid 2,
and $A_{21}(\omega)$ the opposite.

\subsection{Free Fermi Gasses, and Fermi Liquids}

For the sake of comparison, it is worthwhile considering first the
(hypothetical in 1D) situation of coupled free Fermi gasses or Fermi
liquids. For free Fermi gasses, $A_{12}(\omega)\propto\Delta\mu\delta(\omega)$,
$A_{21}(\omega)=0$. Thus $\Gamma(t)\propto\Delta\mu~t$, a clear signal of
coherent hopping and hence of a fundamental rearrangement of the ground
state.

For a true Fermi liquid one finds, using Fermi liquid spectral functions,
\begin{equation}
A_{12}(\omega)\sim\frac{1}{v_F}\theta_+(\omega+\Delta\mu)\{
Z^2\Delta\mu\delta(\omega)+\gamma Z(1-Z)\omega\}
\end{equation}
where $0<Z<1$ is the quasiparticle renormalization factor, and $\gamma$
characterizes the strength of the electron-electron interactions.
$\Gamma(t)$ is therefore a sum of a term $\propto Z^2\Delta\mu t^2$
representing fundamentally coherent processes, and a term
$\propto\gamma Z(1-Z)t^{-1}$ which is on the border of incoherent
and irrelevant. By choosing a sufficiently small $t_{\perp}$ one can
find a time $t$ such that $(1-P(t))/N\ll 1$ ({\em i.e.\/} we are not
outside of the reasonable range of our $O(t_{\perp}^2)$ expansion),
yet the ratio of the coherent contribution to the incoherent contribution
is arbitrarily large. This is true regardless of how small $Z$ is.
Thus, a perturbative calculation in $t_{\perp}$ does not reveal any
likelihood of a loss of coherence of interliquid tunneling, and there is
therefore no impediment to the formation of an interliquid band of
width $\sim Zt_{\perp}$.

\subsection{Luttinger Liquids}

To apply the above method here, we need the electron spectral function
for a Luttinger liquid. In \cite{prl} we used the space-time Luttinger
liquid Green's functions, $G(x,t)$, to calculate $P(t)$, since
these are more directly calculated within the bosonization framework
than are the corresponding $G(k,\omega)$. The electron spectral function
$\rho(k,\omega)$
{\em can\/} be calculated by direct Fourier transform of $G(x,t)$
\cite{spectral}, but
it can actually be determined in a much simpler way \cite{long} by writing the
electron space-time Green's function as a product of ``fracton''
Green's functions, where the fracton operators are exponentials of spin
and charge boson operators. The fracton spectral functions are sharp
$\delta$-functions. By ``glueing'' the fracton spectral functions together
via convolution, one obtains simple integral expressions for the electron
spectral function. For example, for the case of a spin-1/2 Luttinger
liquid with spin-independent interactions (this is the case we are referring to
when we use the term `Luttinger liquid', unless explicitly stated otherwise)
we find
\begin{eqnarray}
{\cal J}_1(k,\omega)
&\propto&
\int_0^{\infty} d\omega_1 d\omega_2 d\omega_3
\delta(\omega-\mu-\sum_i\omega_i)
\delta\left(k-k_F-\frac{(\omega_1-\omega_3)}{v_c}-\frac{\omega_2}{v_s}\right)
\nonumber\\
&&(\omega_1/v_c)^{\alpha-1/2}
(\omega_2/v_s)^{-1/2}
(\omega_3/v_c)^{\alpha-1}
\nonumber\\
\end{eqnarray}
The various singularities near $\omega=\pm v_ck,~v_sk$ can be readily
determined from this expression. $2\alpha$ is the Luttinger liquid
exponent which characterizes the singularity in $n(k)$ near $k_F$ (see
(\ref{n(k)})). For the 1D Hubbard model, $0<2\alpha<1/8$, regardless
of the magnitude of the (on-site) repulsion.

For reasons of pedagogy, it is convenient to first consider the case of
Luttinger liquid models with forward scattering only (FSO), {\em i.e.\/}
where there
is no coupling between left- and right-moving electrons.
Then we shall consider the generic Luttinger liquid case.

\subsubsection{Forward-Scattering-Only Luttinger liquid}

This model exhibits spin-charge separation, but no anomalous exponent
\cite{fabrizio_parola}.
The eigenexcitations are spin and charge bosons with
velocities $v_s$ and $v_c$, respectively, and $v_c-v_s\equiv\Delta v >0$.
We find the rather simple expression
\begin{equation}
A_{12}(\omega)\propto\frac{1}{\Delta v}\theta_+(v_c\Delta k-\Delta\mu-\omega)
\theta_+(\omega+\Delta\mu-v_s\Delta k)
\end{equation}
and $A_{21}(\omega)=0$.

The spectral function for $\Gamma(t)$ is therefore flat $\propto 1/\Delta v$
and non-vanishing only in the region
$v_s\Delta k-\Delta\mu<\omega<v_c\Delta k-\Delta\mu$. It is a simple
matter to explicitly determine $\Gamma(t)$ from $A_{12}(\omega)$, but
the essential physics can be obtained by inspection. In the limit
of $\Delta v\rightarrow 0$, $A_{12}(\omega)\rightarrow\delta(\omega)$.
For $\Delta v\neq 0$, $A_{12}(\omega)$ is peaked around $\omega=0$,
but has a width $\tau^{-1}_{\Delta k}\sim\Delta k\Delta v$. Thus
$\Gamma(t)\propto(\Delta k)t$ for all times $t~^{<}_{\sim}\tau_{\Delta k}$,
which can be made arbitrarily long by choice of sufficiently small $\Delta k$,
while remaining in the perturbative regime, $N^{-1}\int_0^t\Gamma(t')dt'\ll 1$.

We conclude, therefore, that in this case, too, it appears that there is
interliquid coherence for arbitrarily small $t_{\perp}$.

\subsubsection{Generic Luttinger liquid}

We now turn to the case where there is spin-charge separation,
$v_c-v_s=\Delta v > 0$, {\em and\/} an anomalous exponent, $\alpha$.
This case is relevant, for example, to the 1D Hubbard model, and to
most physical models which are not ``chiral''. The expressions for
$A_{12}(\omega)$,
$A_{21}(\omega)$ can be reduced to a one-dimensional integral form \cite{long}.
The key features for $A_{12}(\omega)$ are

(i) a ``spikey'' low-frequency part much like the $A_{12}(\omega)$ in the
FSOLL, above;

(ii) a broad incoherent part, $\propto\omega^{4\alpha}$. More precisely,
for $\omega~^{>}_{\sim}\Delta\mu$
\begin{equation}
A_{12}(\omega)\sim\omega^{4\alpha}+\lambda\Delta\mu\omega^{4\alpha-1}+
O(\Delta\mu)^2
\end{equation}
so that for $\alpha >1/4$ the $O(\Delta\mu)$ contribution is also incoherent.

The resulting $\Gamma(t)$ is as given in \cite{prl}, but with this spectral
representation the picture is somewhat clearer. The potentially coherent
contribution to $\Gamma(t)$ from the low-frequency part of $A_{12}(\omega)$
is $\propto(\Delta\mu)t^{1-4\alpha}$ {\em provided\/}
$t~^{<}_{\sim}\tau_{\Delta k}$;
for $t~^{>}_{\sim}\tau_{\Delta k}$ this term crosses over to {\em incoherent\/}
behavior. The broad background contributes a piece to $\Gamma(t)$ proportional
to
$t^{-4\alpha}$.

Qualitatively, we have the following: there is a (short time) coherent part
with weight $\propto\Delta k$ (or, equivalently, $\Delta\mu$)
coming from the low-frequency part of $A_{12}(\omega)$,
and an incoherent part coming from all but the very lowest frequencies.
Coherence would be favoured by a large coherent contribution, which suggests
making $\Delta k$ large. Increasing $\Delta k$, however, has the effect
of increasing the width, {\em i.e.\/} reducing the ``lifetime'' of the
coherent part. The result is that for arbitrarily small $t_{\perp}$ one
{\em cannot\/} find a time $t$ such that the ratio of the coherent and
incoherent
contributions to $P(t)$ is arbitrarily large. This puts the likelihood of
coherence  in great doubt.

To clarify the physics, let us consider two simple morphologies of
$A_{12}(\omega)$ and their consequences. First, suppose $A_{12}(\omega)$
was the sum of a term $z\delta(\omega)$ and a constant part,
$\gamma~\theta_+(\omega)\theta_+(\Lambda-\omega)$. Then no matter how small $z$
is,
there would always exist a sufficiently small $t_{\perp}$ such that
there was a time $t$, not outside the reasonable range of validity of the
$O(t_{\perp}^2)$ perturbation expansion, for which the coherent contribution to
$P(t)$ is arbitrarily larger than the incoherent contribution. This is true
for the simple reason that the coherent contribution grows as $t^2$,
faster than the incoherent contribution which only grows as $t$.

Now suppose the $\delta(\omega)$ piece is broadened by a lifetime, $\tau$.
Then the contribution to $P(t)$ is now coherent only for $t~^{<}_{\sim}\tau$.
It is then clear that the argument given above for the $z\delta(\omega)$ type
contribution will run into trouble if $z$ is too small, for then
\[
1-P(t)=\delta P_{\rm coh}(t) +\delta P_{\rm incoh}(t)
\]
with
\begin{eqnarray*}
\frac{1}{N}\delta P_{\rm coh}(t)&\sim& zt_{\perp}^2 t^2\;\;\; t~^{<}_{\sim}\tau
\\
&& zt_{\perp}^2 \tau t\;\;\; t~^{>}_{\sim}\tau \\
\frac{1}{N}\delta P_{\rm incoh}(t)&\sim& t_{\perp}^2 \gamma t
\end{eqnarray*}
Clearly, coherence is in doubt if $\gamma~^{>}_{\sim}z\tau$.

We can now see how the Fermi liquid and FSO Luttinger
liquid compare qualitatively to this simple example.
In the former, one effectively has $\tau=\infty$, so the above problem
does not arise. In the latter, while $\tau\sim(\Delta k\Delta v)^{-1}$
there is no incoherent contribution, and is therefore analogous to the simple
model above with $\gamma=0$.

Finally, we point out that in the {\em spinless\/} case $A_{12}(\omega)$
goes over to
\[
A_{12}(\omega)\propto\theta_+(\omega-(v_c\Delta k-\Delta\mu))
(\omega-(v_c\Delta k-\Delta\mu))^{2\alpha-1}
(\omega+(v_c\Delta k+\Delta\mu))^{2\alpha+1}
\]
The presence of a divergent edge singularity for $2\alpha <1$ as
$\omega\rightarrow(v_c\Delta k-\Delta\mu)$ implies that the above
argument for a breakdown of coherence in the spin-1/2 case is not
easily extended to the spinless case.

\section{Conclusion}

We conclude that, within this perturbative approach to the problem
of 1D Luttinger liquids coupled by interliquid hopping, there is
no unambiguous signal of interliquid coherence. This is in contrast
to the case of coupled Fermi liquids and of coupled FSO Luttinger
liquids where a dominant coherent signal emerges at arbitrarily small
interliquid hopping rate, $t_{\perp}$.

Moreover, the $O(t_{\perp}^2)$ result will be modified at higher order
by the presence of interhop ``interactions'' ({\em i.e.\/} correlations).
Such interactions can be expected to favour incoherence, interrupting
the smooth growth of the coherent-like amplitude. An estimate of this
effect was given in \cite{prl}, leading to an approximate expression
for the critical interliquid hopping rate, $t_{\perp}^c$, below which
coherent interliquid hopping will not exist. While the estimate was
crude, the {\em existence\/} of such a $t_{\perp}^c$ would appear beyond
doubt. An extensive discussion of this point will be given in \cite{long}.
The emphasis in this paper is on the absence of an unambiguous signal of
coherence within a simple perturbative ``Golden Rule'' calculation.

We emphasize that, unlike other approaches to this problem \cite{Bunch},
we directly address the coherence/incoherence issue, and
our
method is {\em exact\/}, albeit perturbative. Some other approaches which
claim to go beyond lowest order perturbation theory do so
only at the expense of introducing an uncontrolled approximation.
For example, Boise {\em et al.\/} \cite{boise} use a Wick-theorem resummation
to calculate the single-particle Green's function for coupled liquids. Such
a resummation is invalid, for the $2n$-point correlation functions
in a Luttinger liquid ground state {\em do not\/} satisfy a Wick theorem at
all.
The approximation appears to miss the all-important effects of the incoherent
background.

The application of these ideas to interplanar conduction in the
cuprates, and to understanding the anomalous magnetoresistance in the
quasi-1D organic conductor (TMTSF)$_2$PF$_6$ have been given elsewhere
\cite{prl2,magic}.

We are grateful to P.~W.~Anderson for numerous helpful discussions on this
and related issues.

\newpage

%
%

%
%


\begin{references}
\bibitem {fn_P(t)} Note that `$P(t)$' is used in \cite{TLS}
to refer to $\langle\sigma_z(t)\rangle$. As noted, our $P(t)$
has the same time dependence as $\langle\sigma_z(t)\rangle$, so
the use of different notation is of little consequence. We prefer
to use our definition of $P(t)$ for reasons of unity: in all
sections of this paper, $P(t)$ refers to the probability that the
system, observed at positive time $t$, is found in the same state in which
it was prepared at time $t=0$.
\end{references}
\end{document}